\documentclass{ws-ijmpd}
\usepackage[super,compress]{cite}
 
\usepackage{amssymb,color}
   \usepackage{bm}
\newcommand{\gdual}[1]{\overset{\:{}^{{}^{\boldsymbol{\neg}}}}{\smash[t]{#1}}} 
\newcommand{\dual}[1]{\overset{{}^{{}^{\boldsymbol{\neg}}}}{\smash[t]{#1}}} 
\def\0{\mbox{\boldmath$\displaystyle\mathbb{O}$}}
\def\C{\mbox{\boldmath$\displaystyle\mathbb{C}$}}
\def\R{\mbox{\boldmath$\displaystyle\mathbb{R}$}}

\def\J{\mbox{\boldmath$\displaystyle\boldsymbol{J}$}}
\def\vp{\mbox{\boldmath$\displaystyle\boldsymbol{\varphi}$}}
\def\K{\mbox{\boldmath$\displaystyle\boldsymbol{K}$}}

\def\I{\openone}
\def\s{\mbox{\boldmath$\displaystyle\boldsymbol{\sigma}$}}
\def\p{\mbox{\boldmath$\displaystyle\boldsymbol{p}$}}
\def\e{\rm e}
\def\openone{\mathbb I}

\def\ph{\mbox{\boldmath$\displaystyle\boldsymbol{\hat p}$}}

\begin{document}

\markboth{Dharam Vir Ahluwalia and Alekha Chandra Nayak}
{Elko and mass dimension one field of spin one half: causality and Fermi statistics}

%
\catchline{}{}{}{}{}
%
\title{Elko and mass dimension one field of spin one half: causality and Fermi statistics}

\author{Dharam Vir Ahluwalia$^{a,b}$ and Alekha Chandra Nayak$^a$}

\address{$^a$Department of Physics, Indian Institute of Technology,
Kalyanpur, Kanpur, Uttar Pradesh 208016, India\\
$^b$Department of Physics and Astronomy, University of Canterbury, Christchurch 8140, New Zealand}

\maketitle


\begin{abstract}
We review how Elko arise as an extension of complex valued four-component Majorana spinors. 
 This is followed by a discussion that constrains certain elements of phase freedom. A proof is reviewed that unambiguously establishes that Elko, and for that matter the indicated Majorana spinors, cannot satisfy Dirac equation. They, however do, as they must, satisfy spinorial Klein-Gordon equation. We then introduce a quantum field with Elko as its expansion coefficients and show that it is causal, satisfies Fermi statistics, and then refer to the existing literature to remind that its mass dimensionally is one. We conclude by providing an up-to-date bibliography on the subject. 
\end{abstract}


\ccode{PACS numbers:11.30.Er, 30.70+k,11.10.-z}


\section{Elko}

Elko\footnote{\textbf{E}igenspinoren des \textbf{L}adungs\textbf{ko}njugationsoperators, in German. For higher spins
we will take the liberty of using the acronym Elko to mean `eigen-objects' of the spin-$j$ charge conjugation operator.
This prevents the nomenclature to become too complicated.} are defined as the $(j,0)\oplus(0,j)$ eigen-objects of the spin-$j$ charge conjugation operator. They arose when one of us was at the Los Alamos National Laboratory in the early 1990s, and there was lot of excitement surrounding neutrino oscillations. Understanding their nature, and everything related to neutrinos was simply the inspiration of that exciting time. In order to understand neutrinos he had gotten Majorana's 1937 paper translated into English, and in that context
he
 was studying Pierre Ramond's {\em{primer}}~
[\refcite{Ramond:1981pw}]. In the early chapters Ramond reminds that if $\phi(\p)$ transforms as a massive $(0,1/2)$ Weyl spinor then $\sigma_2 \phi^\ast(\p)$ transforms as a massive $(1/2,0)$ Weyl spinor.  
This observation allows for the introduction of a massive $(1/2,0)\oplus(0,1/2)$ four-component spinor
\begin{equation}
\lambda(\p) \stackrel{\rm def}{=}\left(\begin{array}{cc}
\sigma_2 \phi^\ast(\p)\\
\phi(\p)
\end{array}
\right).\label{eq:one}
\end{equation}
In Ramond's argument the `magic of the Pauli matrices' plays a pivotal role. The interest of the first author at that time was to understand Majorana spinors and construct a higher spin counterpart of $\lambda(\p)$ [nuclear physicists at LAMPF -- the Los Alamos Meson Physics Facilty --  wanted to know how to describe 
higher spin resonances in a systematic manner]. The most natural extension of Ramond's observation where  $\sigma_2 \phi^\ast(\p)$ is replaced by $J_y \phi_{(0,j)}^\ast(\p)$ failed to reproduce the needed magic.\footnote{$J_y$ is the  $(2j+1)\times(2j+1)$ generator of rotation  about the $y$-axis, and $\phi_{(0,j)}(\p)$ is the left-transforming massive Weyl spinor of spin $j$. } The magic, the said author realized, instead  lies in recognizing $\sigma_y = i\Theta$ with $\Theta$ interpreted as spin-1/2 Wigner time reversal operator. Then using the property that for any spin, the Wigner time reversal operator satisfies 
\begin{equation}\Theta_j \J \Theta_j^{-1} = -\J^\ast
\end{equation}
allows the needed magic to happen, that is: if $\phi_{(0,j)}(\p)$ is a left-transforming massive spin-$j$ Weyl spinor belonging to the $(0,j)$ representation space, then $\zeta \Theta 
\phi^\ast_{(0,j)}(\p)$, with $\zeta \in \C \vert \zeta^\ast\zeta =1$, transforms as 
a right-handed massive spin-$j$ Weyl spinor belonging to the $(j,0)$ representation space.
The demand that ${(j,0)\oplus(0,j)}$ extension  of (\ref{eq:one})
\begin{align}
\lambda_j(\p) \stackrel{\rm def}{=} 
\left(\begin{array}{cc}
\zeta\Theta_j\phi_j ^\ast(\p)\\
\phi_j(\p)
\end{array}
\right)\label{eq:three}
\end{align}
be an eigen-object of the spin-$j$ charge conjugation operator then determines the phase $\zeta$. 
\vspace{11pt}

For $j=1/2$, one finds that $\zeta = \pm i$ and
\begin{equation}
\Theta = \left(\begin{array}{cc}
0 & -1 \\
1 & 0
\end{array}
\right).
\end{equation}
For spin one half, one thus has not  two, but four, Elko.  The self-conjugate Elko are the standard Majorana spinors interpreted as complex-valued four-component spinors. While the existence of the other two Elko, no longer allows the set of four Elko to be interpreted as Weyl spinors in disguise. As far as spinors are concerned they are as fundamental as the well-known Dirac spinors: Elko are the eigenspinors of the charge conjugation operator while the Dirac spinors are the eigenspinors of the Parity operator.

While the 1937 Majorana field can be written in  a form that 
resembles $\lambda_{j=1/2}(\p)$ it is to be recalled that the field is still expanded in terms of the Dirac spinors, and not Elko.\footnote{When no confusion is likely to arise
we will simply write $\lambda_{j=1/2}(\p)$ as $\lambda(\p)$.}
 This superficial resemblance, if not properly clarified, can be a cause of much confusion. In fact the 1937 paper is a simple statement of the fact that if one sets $b_\sigma^\dagger(\p) = a_\sigma^\dagger(\p)$ in the Dirac field then the charge carried by such a field is identically equal to zero. The 1937 field has nothing directly to do with Majorana spinors. To our knowledge, these arose only some two decades later in an attempt to reformulate Majorana field~[\refcite{McLennan:1957JA,Case:1957zza}].
 
 \section{Phases in Elko}\label{Sec:Phases}
 
For an arbitrary spin, there are $(2j+1)$ self-conjugate Elko, and an equal number of anti-selfconjugate Elko. Regarding $\phi_j(\p)$ that appear in $\lambda_j(\p)$,
beyond the fact that they transform as left-handed spin-$j$ Weyl objects,
the Elko defining arguments place no other constraints on  $\phi_j(\p)$. If a choice of phases is made for the $\phi_j(\p)$, then this introduces a relative phase between the right- and left-transforming components of $\lambda_j(\p)$. A specific choice of these phases affects the physical and mathematical properties of spin-$j$ Elko. While there is no a priori reason to take $\phi_j(\p)$ as eigen-objects of $\J\cdot\ph$, this is the simplest choice that has been explored so far.  One reason for this is that the same operator also enters in the rotation and boost transformations of the $(j,0)\oplus(0,j)$ objects. The generators for the latter, $\K$, equal $\pm i$ times the generators for the former, $\J$. 
\vspace{11pt}

While a general freedom of  a global phase factor is not permitted for Elko, a restricted set of phases equal to $\pm 1$ are allowed for each of the Elko without destroying their self/anti-selfconjugacy under the charge conjugation operator. This freedom, and the choice of the phases mentioned above, dramatically affects the locality anticommutators of the mass dimension one fermions of spin one half and their extensions to higher spins~[\refcite{Lee:2012td}].
\vspace{11pt}

The 2005 publications mentioned above did not take these phases into account~[\refcite{Ahluwalia:2004sz,Ahluwalia:2004ab}]. The full appreciation of these phases emerged in a subsequent series of papers~[\refcite{Ahluwalia:2008xi,Ahluwalia:2010zn}]. They, in part, determine the locality structure of the originally-reported quantum field. The other part that enters the locality story is an observation that we will soon make below. But before we enter that thread let us show that the Dirac operator does not annihilate Elko.

\section{Elko and the Dirac operator}

The canonical spin one half Elko are obtained by choosing in equation (\ref{eq:one}), $\phi(k^\mu)$ as eigenspinors of the helicity operator $(1/2)\s\cdot \ph$, 
\begin{equation}
\s\cdot\ph \;\phi_\pm(k^\mu) = \pm \phi_\pm (k^\mu)
\end{equation}
where 
\begin{equation}
k^\mu =\left(\begin{array}{l}
m \\
\frac{\p}{p}\Big\vert_{p\to 0}
\end{array}\right)
\end{equation} 
The phases associated with the Weyl spinors at `rest' are then chosen as in Ref.~[\refcite{Ahluwalia:2008xi,Ahluwalia:2009rh}]. The arbitrary-momentum Elko now follow by boosting the $\lambda(k^\mu)$
\begin{align}
\lambda(\p) &= \left(
\begin{array}{cc}
\exp(\frac{1}{2}\s\cdot\vp) & \0 \\
\0 & \exp(- \frac{1}{2}\s\cdot\vp)
\end{array}
\right) \lambda(k^\mu) \nonumber\\
&= 
\sqrt{\frac{E+m}{2 m}}
\left(\begin{array}{cc}
\I + \frac{\s\cdot \p}{E+m} & \0 \\
\0 & \I - \frac{\s\cdot \p}{E+m} 
\end{array}
\right)\lambda(k^\mu).\label{eq:seven}
\end{align}
This exercise yields a set of four Elko, two of them are self conjugate $\lambda^S_\pm(\p)$, and the other two, $\lambda^A_\pm(\p)$, are anti-selfconjugate under the action of the charge conjugation operator. The action of the Dirac operator on Elko can then be readily evaluated by using two  observations, (a) the Elko at rest are
\begin{equation}\lambda^S_\pm(k^\mu) = \left(
\begin{array}{c}
i \Theta \phi^\ast_\pm(k^\mu)\\
\phi_\pm(k^\mu)
\end{array}
\right), \quad
\lambda^A_\pm(k^\mu) = \pm \left(
\begin{array}{c}
- i \Theta \phi^\ast_\mp(k^\mu)\\
\phi_\mp(k^\mu)
\end{array}
\right)
\end{equation}
with phases for $\phi_\pm(\p)$ chosen as in Ref.~[\refcite{Ahluwalia:2008xi,Ahluwalia:2009rh}], and (b)
that the helicity of $i \Theta \phi^\ast_\pm(k^\mu)$ is opposite to that of $\phi_\pm(k^\mu)$. Combined with equation (\ref{eq:seven}
) this yields
\begin{align}
\lambda^S_\pm (p^\mu) = \sqrt{\frac{E+m}{2 m}}\left(1 \mp \frac{p}{E+m}\right)\lambda^S_\pm(k^\mu)\\
\lambda^A_\pm (p^\mu) = \sqrt{\frac{E+m}{2 m}}\left(1 \pm \frac{p}{E+m}\right)\lambda^A_\pm(k^\mu).
\end{align}
\vspace{11pt} 

\noindent
Now we can proceed further, and explicitly note that
\begin{equation}
\gamma_\mu p^\mu \lambda^S_+(p^\mu) =
\left[ E\gamma_0 + 
p\left(\begin{array}{cc}
\0 & \s\cdot\p \\
-\s\cdot\p &\0
\end{array}
\right)
\right] \sqrt{\frac{E+m}{2 m}}\left(1 - \frac{p}{E+m}\right)\lambda^S_+(k^\mu).
\end{equation}
But since
\begin{align}
\left(\begin{array}{cc}
\0 & \s\cdot\p \\
-\s\cdot\p &\0
\end{array}
\right)
&\lambda^S_+(k^\mu)  = \gamma_0 \lambda^S_+(k^\mu) 
\end{align}
we have
\begin{equation}
\gamma_\mu p^\mu \lambda^S_+(p^\mu) =
\left( E + p \right)
 \sqrt{\frac{E+m}{2 m}}\left(1 - \frac{p}{E+m}\right)\gamma_0\lambda^S_+(k^\mu).
\end{equation}
Next we use the identities
\begin{align}
& \gamma_0 \lambda^S_+(k^\mu)  = i \lambda^S_-(k^\mu)
\\
 (E+p) \sqrt{\frac{E+m}{2 m}}\left(1 - \frac{p}{E+m}\right) & =
\sqrt{\frac{E+m}{2 m}}\left(1 + \frac{p}{E+m}\right) \times m
\end{align}
and 
\begin{equation}
\sqrt{\frac{E+m}{2 m}}\left(1 + \frac{p}{E+m}\right)\lambda^S_\pm(k^\mu) = \lambda^S_-(p^\mu)
\end{equation}
to arrive at 
\begin{equation}
\gamma_\mu p^\mu \lambda^S_+(p^\mu) = i m \lambda^S_-(p^\mu).
\end{equation}
Since similar results hold for the remaining three Elko, the above equation and the remaining three can be combined into the following four equations
\begin{equation}
\gamma_\mu p^\mu \lambda^{S,A}_\pm (p^\mu) = \pm i m \lambda^{S,A}_\mp(p^\mu).
\label{eq:notdirac}
\end{equation}
and arrive at the conclusion that Elko cannot satisfy the Dirac equation.\footnote{This result was first arrived at by Valeri Dvoeglazov, see references~\refcite{Dvoeglazov:1995eg,Dvoeglazov:1995kn}. However, the derivation presented here is original and significantly more transparent.}
\vspace{11pt} 

Equations~(\ref{eq:notdirac}) immediately combine to show that Elko do satisfy the spinorial Klein-Gordon equation
\begin{equation}
(\eta_{\mu\nu}p^\mu  p^\nu \I_4 - m^2\I_4)\lambda(p^\mu) = 0, \label{eq:skg}
\end{equation}
where $\eta_{\mu\nu}={\rm diag}\{1,-1,-1,-1\}$.

\section{From Elko to Mass dimension one field: causality and fermionic statistics}

Since under the Dirac dual $\overline\lambda(p^\mu) \stackrel{\rm def}{=} \lambda^\dagger(p^\mu) \gamma_0$, the scalar product $\overline\lambda(p^\mu) \lambda(p^\mu)$ identically vanishes,
 the route from here to the construction of mass dimension one fermionic fields  requires the introduction of a new spinorial dual $\dual\lambda(p^\mu)$  in the $(1/2,0)\oplus(0,1/2)$ representation space. It can be found in reference~[\refcite{Ahluwalia:2008xi,Ahluwalia:2009rh}].
 Once that is done one may calculate the Elko spin sums, and introduce a quantum field, $\mathfrak{f}(x)$,  with $\lambda^{S,A}_\pm(p^\mu)$ as its expansion coefficients
\begin{equation}
\mathfrak{f}(x)  =   \int \frac{{d}^3p}{(2\pi)^3}  \frac{1}{\sqrt{2 m E(\p)}} \sum_\alpha \Big[  a_\alpha(\p)\lambda^S_\alpha(\p) \e^{-i p\cdot x}
+\, b^\dagger_\alpha(\p)\lambda^A_\alpha(\p) \e^{i p\cdot x}{\Big]}
\label{eq:mdof}
\end{equation}
and its associated adjoint
\begin{align}
\gdual{\mathfrak{f}}(x) 
  =   \int \frac{{d}^3p}{(2\pi)^3}  \frac{1}{\sqrt{2 m E(\p)}}  \sum_\alpha   \Big[ a^\dagger_\alpha(\p) \dual{\lambda}^S_\alpha(\p) \e^{i p\cdot x}  + b_\alpha(\p)\dual{\lambda}^A_\alpha(\p) \e^{-i p\cdot x}{\Big]}. \label{eq:fadj}
\end{align}
In the definition of $\mathfrak{f}(x)$ and $\gdual{\mathfrak{f}}(x) $ the statistics of the creation and destruction operators is left undetermined. However, the phases mentioned in section \ref{Sec:Phases} must not be ignored.
\vspace{11pt}

To settle the statistics, consider two \emph{space-like} separated events, $x$ and $y$. For space-like separations temporal ordering of events is not necessarily preserved, and thus there exist two sets of inertial frame, ones in which $y^0>x^0$ and the ones in which the reverse is true, $x^0>y^0$. We call these sets of inertial frame as $\mathcal{O}$ and  $\mathcal{O}^\prime$ respectively. 
In $\mathcal{O}$,
we calculate the amplitude for a particle to propagate from $x$ to a space-like separated point $y$ and in $\mathcal{O}^\prime$  the amplitude for an antiparticle to propagate from the point $y$ to the space-like separated point $x$.\footnote{For doing the calculations we go  from $\mathcal{O}$ to $\mathcal{O}^\prime$ 
with a Lorentz transformation that interchanges $x\leftrightarrow y$: $(x-y) \to -(x-y)$.}
 Causality requires
 these two amplitudes may only differ by a phase:\footnote{The essence of this argument can be traced back to Feynman~[\refcite{Feynman:1986rp}] and Weinberg~[\refcite[Sec.~2.13]{Weinberg:1972ss}]. It differs from a later argument presented on the same subject by Weinberg in Ref.~[\refcite{Weinberg:1995mt}] where the pivotal role is played by the cluster decomposition principle.} 
\begin{equation} 
{\rm Amp}(x\to y, {\rm particle})\big\vert_\mathcal{O} =  e^{i \theta} {\rm Amp}(y\to x, {\rm antiparticle})\big\vert_{\mathcal{O}^\prime} \label{eq:amp}
\end{equation}
with $\theta\in\R$.
The calculated Elko spin sums force these two amplitudes to have opposite 
signs.\footnote{In order not to make our notation `heavy' we are using the same symbols $x$ and $y$, both in $\mathcal{O}$ and $\mathcal{O}^\prime$ to mark the events under consideration. The distinction is then made unambiguous  by explicitly referring the said frames.}  
 Outline of this result is as follows
 \begin{align}
 {\rm Amp}(x\to y, {\rm particle})\big\vert_\mathcal{O} &= \langle\hspace{5pt}\vert
\mathfrak{f}(y)\gdual{\mathfrak{f}}(x) \vert\hspace{5pt}\rangle\nonumber\\
{\rm Amp}(y\to x, {\rm antiparticle})\big\vert_{\mathcal{O}^\prime} & = 
\left[
 \langle\hspace{5pt}\vert
\gdual{\mathfrak{f}}(x){\mathfrak{f}}(y) 
\vert\hspace{5pt}\rangle\big\vert_{\mathcal{O}}\right]\Big\vert_{(x-y)\to (y-x)} \nonumber\\
& = - {\rm Amp}(x\to y, {\rm particle})\big\vert_\mathcal{O}\label{eq:25}
\end{align}
Substituting ${\rm Amp}(y\to x, {\rm antiparticle})\big\vert_{\mathcal{O}^\prime} $ from (\ref{eq:25}) into (\ref{eq:amp}), gives $e^{i\theta} = -1$.
The ${\rm Amp}(y\to x, {\rm antiparticle})\big\vert_{\mathcal{O}^\prime} $ and 
${\rm Amp}(y\to x, {\rm antiparticle})\big\vert_{\mathcal{O}}$ must yield the same probability, and thus they can differ by a phase alone:
\begin{equation}
{\rm Amp}(y\to x, {\rm antiparticle})\big\vert_{\mathcal{O}^\prime} 
= e^{i\tilde\theta}
{\rm Amp}(y\to x, {\rm antiparticle})\big\vert_{\mathcal{O}} \label{eq:zimpok11}
\end{equation}
with $\tilde\theta\in\R$.
However, one can go a step further and explicitly calculate 
the two indicated amplitudes in the above equation. We find them to be same. That 
is $e^{i\tilde\theta} =1$. Using this result, (\ref{eq:amp}) becomes:
\begin{equation}
{\rm Amp}(x\to y, {\rm particle})\big\vert_\mathcal{O} + {\rm Amp}(y\to x, {\rm antiparticle})\big\vert_{\mathcal{O}} = 0. \label{eq:amp-c}
\end{equation}
That is, $\mathfrak{f}(x)$ is fermonic
\begin{equation}
\langle\hspace{5pt}\vert\{\mathfrak{f}(x),\gdual{\mathfrak{f}}(y)\}\vert\hspace{5pt}\rangle = 0,\label{eq:stat}
\end{equation}
and that the amplitude for  a particle 
to propagate from $x$ to $y$ and and an antiparticle to propagate from $y$ to $x$ in a given inertial frame are opposite.\footnote{In all the above calculations we did not use the above arrived anti-commutativity.}  
\vspace{11pt}

To decipher the mass dimensionality associated with the set $\mathfrak{f}(x)$ and $\gdual{\mathfrak{f}}(x)$ we calculate 
the Feynman-Dyson propagator,
$S_{FD}(x - x^\prime)  = i \langle\hspace{5pt}\vert\mathfrak{T}[\mathfrak{f}(x^\prime)\gdual{\mathfrak{f}}(x)]\vert\hspace{5pt}\rangle$. 
Exploiting the just derived Fermi statistics, yields
\begin{align}
S_{FD}(x - x^\prime)  
& = i \left[ {\rm Amp}(x\to x^\prime, {\rm particle})\big\vert_{t^\prime>t}
\textcolor{black}{-} {\rm Amp}(x^\prime\to x, {\rm antiparticle})\big\vert_{t>t^\prime}
\right]\nonumber\\
& = i \left[\langle\hspace{5pt}\vert
\mathfrak{f}(x^\prime)\gdual{\mathfrak{f}}(x) \vert\hspace{5pt}\rangle\theta(t^\prime - t)
-  \langle\hspace{5pt}\vert
\gdual{\mathfrak{f}}(x){\mathfrak{f}}(x^\prime) 
\vert\hspace{5pt}\rangle\theta(t-t^\prime)
\right].\label{eq:fdpropagator}
\end{align}
In each of the two terms the indicated time orderings can be split into space-like and non space-like events. Because of the observation made following equation~(\ref{eq:stat}) the contributions for the former add. In consequence, $S_{FD}(x - x^\prime)$ allows propagation outside the light cone.
\vspace{11pt}

We have thus derived, rather than assumed, Fermi statistics for $\mathfrak{f}(x)$ and  $\gdual{\mathfrak{f}}(x)$. 
The $S_{FD}(x - x^\prime)$ has already been evaluated in the existing literature but its form now  stands on firmer theoretical grounds. It endows  field $\mathfrak{f}(x)$ with mass 
dimension 
one~[\refcite{Ahluwalia:2004sz}]-[\refcite{Ahluwalia:2008xi}].\footnote{The last two references incorporate the phases mentioned in section
\ref{Sec:Phases}. These phases do not affect the mass dimensionality, but yield a dramatically improved locality structure.}

\section{Bibliography: Elko and Mass Dimension One Fermions}

Elko and mass dimension one fermions are now an active area of research. A brief review of these investigations has been provided in the Foreword. Since the form of the latter does not allow a detailed bibliography we here list the relevant works~[\refcite{daRocha:2005ti}] 
to~[\refcite{Jardim:2014xla}].


\section*{Acknowledgments}

It is my (DVA) pleasure to thank all my collaborators and many friends in the Elko community. I owe them immense gratitude for so many discussions, hospitality, and for freely sharing their ideas.

\end{document}